# A Curated Rotamer Library for Common Post-Translational Modifications of Proteins

Oufan Zhang[1], Shubhankar A. Naik[2], Zi Hao Liu[5,6], Julie Forman-Kay[5,6], Teresa Head-Gordon[1-4]*

[1]Kenneth S. Pitzer Center for Theoretical Chemistry, [2]Department of Chemistry, [3]Department of Bioengineering, and [4]Department of Chemical and Biomolecular Engineering, University of California, Berkeley, Berkeley, California 94720, USA

[5]Molecular Medicine Program, Hospital for Sick Children, Toronto, Ontario M5G 0A4, Canada, [6]Department of Biochemistry, University of Toronto, Toronto, Ontario M5S 1A8, Canada

Sidechain rotamer libraries of the common amino acids of a protein are useful for folded protein structure determination and for generating ensembles of intrinsically disordered proteins (IDPs). However much of protein function is modulated beyond the translated sequence through thFiguree introduction of post-translational modifications (PTMs). In this work we have provided a curated set of side chain rotamers for the most common PTMs derived from the RCSB PDB database, including phosphorylated, methylated, and acetylated sidechains. Our rotamer libraries improve upon existing methods such as SIDEpro and Rosetta in predicting the experimental structures for PTMs in folded proteins. In addition, we showcase our PTM libraries in full use by generating ensembles with the Monte Carlo Side Chain Entropy (MCSCE) for folded proteins, and combining MCSCE with the Local Disordered Region Sampling algorithms within IDPConformerGenerator for proteins with intrinsically disordered regions.

*Corresponding authors: thg@berkeley.edu

# INTRODUCTION

Post-translational modifications (PTMs) refer to the chemical modifications that are made to the amino acids of a protein after translation to enable the cell to regulate its function. The importance of PTMs can't be understated given that at least 80% of mammalian proteins are modulated by PTMs, influencing essentially all biological processes, including cell signaling and metabolic pathways, transcriptional regulation, and DNA repair. In addition, dysregulation of PTMs is implicated in the development and progression of many diseases including cancer.[1] In spite of the fact that there are hundreds of PTMs known to occur in biology,[2] the available experimental structural data, e.g. from X-ray crystallography, cryo-electron microscopy (Cryo-EM) or Nuclear Magnetic Resonance (NMR) measurements, remains sparse across the full space of chemical modifications compared to unmodified proteins.

In principle, computational models can fill the gap for modeling the protein structural changes introduced by PTMs, although most structure-based prediction algorithms are primarily designed for the canonical amino acids. For example, while the original AlphaFold2[3] and RosettaFold[4] have revolutionized protein structure prediction for unmodified sequences, they did not handle PTMs. However, these algorithms are also largely limited to prediction of single stable protein conformations, and do not provide insights into protein flexibility[5-7], including the most flexible class with intrinsic disorder.[8-12] PTMs modulate protein energy landscapes, which can lead to changes in conformations and in their dynamic interconversions. Protein flexibility also leads to the fluctuations of side chain packing arrangements that often have a direct functional role[13-17]. A large number of high-quality physical algorithms[18-23] and machine learning approaches[24-26] exist to perform side chain repacking for the canonical amino acids for folded proteins, disordered proteins when they undergo binding-upon-folding, or when they form dynamical complexes.

However, the ability to model side chain ensembles with PTMs are currently quite limited. There have been a small number of studies for computational modeling of PTMs[27]; software packages such as Rosetta[28,29], FoldX[30], and SIDEpro[25] have the ability to model PTMs using rotamer libraries[31]. A recent method named GlycoSHIELD exclusively focuses on the modeling of glycosylation by grafting glycan conformer candidates sampled from molecular dynamics trajectories[33]. Given that experimental PTM structural data has continued to accumulate since many of these methods have been developed, it is worth creating new side chain rotamer libraries containing PTMs. Furthermore, since some rotamer libraries such as SIDEpro only characterized the effect of the PTM modifications on side chain torsion angles, it is also valuable to account for the rotamer distribution shifts of the backbone dihedral angles.

In this study, we have undertaken the creation of both backbone-dependent and backbone-independent rotamer libraries to comprehensively investigate the influence of PTMs on sidechain conformational ensembles. The decision to generate both types of libraries arises from the goal of capturing nuanced details of sidechain variability within the local structural context provided by the protein's backbone, as well as to understand more general patterns of sidechain conformations across diverse protein structures where data sparsity is less of a concern. We compare our rotamer libraries against SIDEpro[25] and Rosetta[28,29], and show that the overall trend across various metrics show systematic improvements against folded proteins and IDP compared to these standard methods.

# METHODS

*Structural Data for post-translational modifications of amino acids.* To generate structured data for PTM-modified amino acids to construct our datasets, we first accessed the PTM Structural Database (PTM-SD).[32] Each data entry in the PTM-SD corresponds to a distinct post-translational modification on a specific amino acid residue. Within the PTM Structural Database, an inquiry targeting these amino acids and their associated modifications facilitated the retrieval of RCSB Protein Data Bank (PDB) identification codes for each modified residue.

Utilizing the obtained PDB identification codes, we conducted searches within the PDB database[33] to acquire structural data for the corresponding modified amino acids. A notable observation was that a subset of proteins frequently exhibited high sequence similarity and identity. Such a high degree of redundancy would introduce biases that could impact the interpretation of patterns and relationships in the rotamer data. To address this, the independent NCBI BLAST tool[34] was employed to detect identical sequences. Through the alignment of FASTA sequences from all proteins within the dataset, a threshold for sequence similarity of 90% or above was established, resulting in the clustering of protein structures. Subsequently, within each group, the structure having the highest resolution was selected for further analysis. Furthermore, PDB files with incomplete structural data were excluded.

Supplementary Tables S1 and S2 provide the curated PDB dataset for PTM-containing amino acids, further delineated by the refined resolution ranges for structures including each modified residue type, and the number of PDB files available in each resolution category. We found PDB files for phosphorylated serine (SEP), phosphorylated threonine (TPO), phosphorylated tyrosine (PTR), methylated arginine (AGM), mono-methylated lysine (MLZ), di-methylated lysine (MLY), tri-methylated lysine (M3L), oxidized methionine (OMT), and acetylated lysine (ALY) (see Supplementary Table S3). However, we chose to perform our analysis and rotamer generation to create PTM rotamer libraries only on modified amino acids having sufficient data by defining a resolution cutoff of 3.5 Å and lower, and requiring a minimum of 40 total PDB files. Based on these criteria, we only developed rotamer libraries for SEP, TPO, PTR, M3L, and ALY.

*Statistical Analysis for PTM rotamer libraries*. Given that the modified amino acids exhibit a non-uniform distribution of backbone phi ($\phi$) and psi ($\psi$), and the relative sparsity of the observations of side chain rotamer chi ($\chi_i$) for specific backbone dihedral angles, we use the adaptive kernel density estimation of Shapovalov and Dunbrack to estimate the rotamer probability density functions (PDFs)[35]. This method determines the width of the kernel based on the local density of data points, such that in denser regions narrower kernels are applied to capture more local variations while broader kernels create smoother distributions in regions where rotamer occupancy is sparse.

We employ the von Mises kernel[36] which is well-suited to periodic data such as dihedral angles. The Nadaraya-Watson kernel regression model[37,38] considers the influence of neighboring data points in a weighted manner, producing smoothed estimates of the mean and standard deviation values for $\chi$ angles within each $\phi/\psi$ bin. This smoothing helps mitigate the impact of noise and fluctuations in the data, providing a more robust and reliable characterization of the relationships between the backbone $\phi/\psi$ and sidechain $\chi$ dihedral angles. Bayes' rule is applied to these rotamer probability density functions to acquire the rotamer probabilities.

*Backbone-dependent (BD) PTM rotamer libraries*. In the construction of the BD-rotamer library, we discretized the backbone space into bins, each spanning a 30° interval. This level of granularity allows for a detailed representation of the local backbone geometry and facilitates the accurate prediction of sidechain conformations. The 30° bin size was chosen to balance computational efficiency with the need to capture subtle variations in sidechain orientations influenced by the local backbone structure. Regarding discretized bins which lack occupancy in the rotamer library, the probability and $\chi$ angle means and standard deviation are estimated using the backbone-independent (BI) probability distributions.

In the process of estimating rotamer means and standard deviations and discretizing angle bins for a rotamer sidechain library, the specified angle ranges for trans (T, (120°, 180°) or (-180°, -120°)), gauche (G$^+$, (0°, 120°)), and gauche- (G$^–$, (-120°, 0°)) conformers are defined for sp$^3$ carbons (Supplementary Figure S1). Given a sp$^3$-sp$^3$ hybridized bond, the degrees of freedom of the dihedral angles have probability density distributions that contain 3 distinct and symmetric peaks that occur at 60°, -60°, and -180/180° and that align with the conformers mentioned above. For ALY $\chi_5$, the rotamer bins are instead defined for gauche (G$^+$, (30°, 90°) or (-150°, -90°)), gauche- (G$^-$, (90°, 150°) or (-90°, -30°)) and trans (T, (-30°, 30°) or (-180°, -135°) or (135°, 180°)) given the sp$^3$-sp$^2$ hybridized bond. For PTR $\chi_2$, the rotamer bins are defined for gauche (G, (45°, 135°) or (-135°, -45°)) and trans (T, (-45°, 45°) or (-180°, -135°) or (135°, 180°)) to account for the symmetry of a benzene ring. When discretizing angle bins, they are aligned with these conformer-specific ranges, ensuring that each bin corresponds to the appropriate conformer type. The estimation of rotamer means and standard deviations is then performed within these conformer-specific ranges.

While canonical amino acids fit well into this discretization strategy, we find that PTMs sometimes do not fit into standard categories. To accommodate non-standard rotamer angle distributions observed for the $\chi_2$ of SEP and TPO, and $\chi_3$ of PTR, we split the probability distributions by their population density with clusters defined using DBSCAN[43]. Furthermore, as the terminal $\chi$ angles of phosphorylated SEP, TPO, PTR, and the methylated M3L have a 3-fold symmetry around the torus axis and show broad distributions regardless of the other sidechain angles, we estimated their means and standard deviations by only aligning with the conformer ranges of the angles themselves (Supplementary Figure 2).

*Creating protein structures using PTM rotamer libraries for folded proteins and IDPs*. We use our newly generated PTM libraries by generating side chain ensembles for folded proteins with the Monte Carlo Side Chain Entropy (MCSCE).[18] The MCSCE algorithm is also embedded within the Local Disordered Region Sampling (LDRS)[39] algorithms within IDPConformerGenerator[40] for proteins with intrinsically disordered regions (IDRs). The backbone conformers of disordered regions were aligned and attached to folded domains using the LDRS module in IDPConformerGenerator, discarding conformations with steric clashes. For each disordered region case, we sampled 20,000 backbone conformations from a combination of loop, helices, and β-strands regions; successful and complete sidechain packing statistics are given in Supplementary Table S4.

Candidate sidechain packings were generated with Monte Carlo-Side Chain Entropy (MCSCE)[18] program for the disordered regions of the unmodified sequence using the Dunbrack library[39,44], and with PTMs using the BD-rotamer and BI-rotamer libraries we developed in this work. We also used MCSCE to generate sidechain packings using the SIDEpro[25] PTM rotamer libraries. As the SIDEpro rotamer library only defines the additional sidechain angles for PTMs conditioned on the rotamer ranges of the existing angles in the unmodified amino acid residues, these sidechain angles were sampled according to the probability distribution of the corresponding canonical amino acids. In both cases, the rotamers of the folded domains are unchanged. The Rosetta[28,29] packed conformers were generated by invoking the fixbb application starting from the same backbone conformers attached to the folded domains, similarly with the folded domains held fixed. As Rosetta's scoring function during conformer packing does not define an explicit clash cutoff as in MCSCE, the Rosetta generated conformers were also filtered with the same clash criteria in MCSCE for comparison.

## RESULTS

We first consider a BD-rotamer library for the SEP, TPO, PTR, M3L, and ALY that categorize sidechain conformations based on specific backbone $\phi$ and $\psi$ dihedral angles. As shown by Dunbrack and co-workers[41], a BD library for predicting side chain conformations produces much better results for protein structure refinement using NMR and X-ray data as opposed to a BI-rotamer library. The BI-rotamer library focuses solely on the distribution of sidechain dihedral angles, and for PTMs may be a necessity if the data is too sparse to differentiate it from the BD-rotamer case.

The influence of backbone conformations on PTM sidechain rotamer states is illustrated in Figures 1 and 2 (and Supplementary Figures S1-S3). In Figure 1, the $\chi_1$ of PTM-modified amino acids show distinct rotamer populations in different $\phi/\psi$ regions (and in relation to secondary structure), and more importantly the $\chi_1$ distributions for the PTM-modified amino acids also shift considerably from the unmodified canonical residues in these regions. For example, while the $\chi_1$ of SER, THR and LYS show similar populations regardless of $\phi/\psi$ ranges, the $\chi_1$ of SEP, TPO and M3L have visibly different preferences for certain backbone regions (Figure 1). This illustrates that the SIDEPro rotamer library that utilizes an unmodified $\chi_1$ will be unable to fully capture these structural changes exhibited by the PTMs.

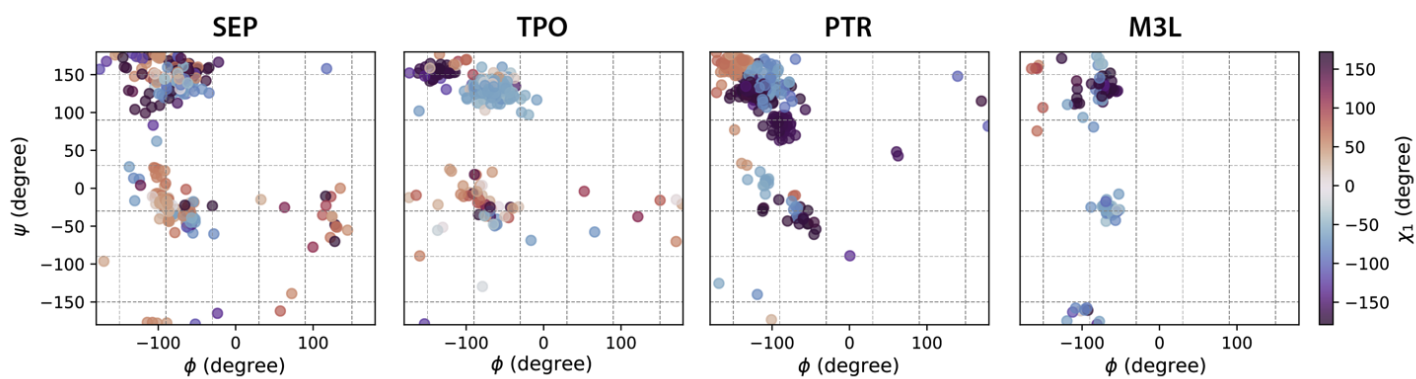

**Figure 1**. *Ramachandran plots color coded by $\chi_1$ angle ranges for PTM-modified amino acids using the backbone-dependent library.* Backbone bins in 60° separation are shown in gray dash lines.

The $\chi_2$ of the phosphorylated amino acids SEP and TPO adopt rotamers that cannot be easily explained using bond hybridization models, but instead arise from favorable hydrogen bonding interactions[45] (Figure 2). In particular, TPO has a dominating $\chi_1/\chi_2$ population centered at -60°/120°, a configuration that encourages formation of an internal P-O/N-H hydrogen bond (Figure 2B). $\chi_2$ of SEP exhibits a mixed population that spans the angular range from 60° to -120°. In addition to the $\chi_1/\chi_2$ -60°/±120° configuration associated with an internal hydrogen bond (Figure 2-A1), we observe the formation of a hydrogen bond between P-O of the SEP residue *i* and N-H of its adjacent residue *i+1* with a $\chi_1/\chi_2$ configuration around -180°/60° (Figure 2-A2). These varieties of sidechain rotamer states are consistent with the cooperative transition between a state in which the phosphate group is well-solvated and a state that forms intra- and inter-residue hydrogen bonds, as noted in reference (45). As shown in Figure 2 (and Supplementary Figures S1-S3), the observed higher percentage of hydrogen bond formations in the polyproline helical backbone region, and the resulting selectivity in the $\chi_1/\chi_2$ configurations, help rationalize the distinct rotamer populations for phosphorylated amino acids such as SEP and TPO that are dependent on backbone configurations. All these differences highlight the need to better characterize the rotamer states of PTM-modified amino acids apart from the canonical amino acids.

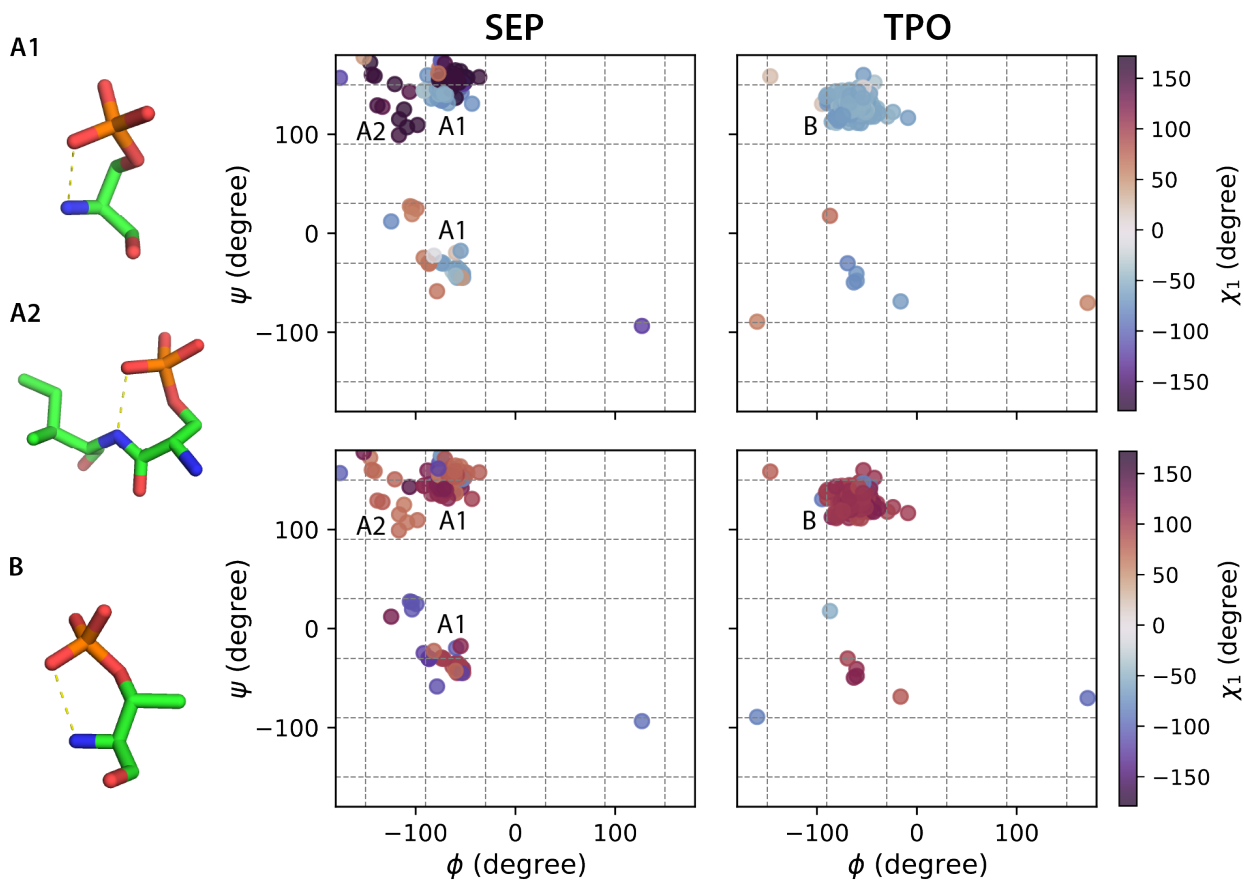

**Figure 2**. *Ramachandran plots color coded by $\chi_1$ and $\chi_2$ angle ranges for SEP and TPO with hydrogen bond formation using the backbone-dependent library.* Backbone bins in 60° separation are shown in gray dash lines. Hydrogen bonds are defined by within a donor-acceptor distance cutoff of 3.5 Å. A1, A2 and B show representative configurations for SEP (A1, A2) and TPO (B) associated with the annotated backbone regions.

We verify the fidelity of our constructed BD-rotamer and BI-rotamer libraries for PTMs by sampling their sidechain torsion distributions to generate new side chain packings and compare the resulting accuracies of the repacked structures to the experimental PDB data. In Supplementary Figure S4, we show that the sidechain rotamer distribution of the constructed libraries are in good agreement with the PDB data when sampled based on the same backbone conformation. When repacking PTMs sidechains, we removed the original PTM sidechain structures and then resampled the rotamers for PTMs with all other residues intact. We compare the RMSD values of the full protein structures with PTMs, regenerated with the BD-rotamer and BI-rotamer libraries from this work against the SIDEpro and Rosetta libraries, and comapre the distributions as boxplots in Figure 3A.

It is evident that the BD rotamer library has a smaller interquartile range of somewhere between 0.25 Å to 1.0 Å and is skewed toward lower RMSD values across all of the PTM types compared to the other rotamer libraries (Figure 3A and Supplementary Table S5). The BI-rotamer library for PTMs also demonstrates trends in improvement to the other standard methods, with the exception of TPO, and the statistical significance of improvement is not as strong compared to the BD-rotamer library. We also performed a Wilcoxon Signed Rank test to evaluate the RMSD differences between each distribution pair with a 95% confidence level, in which the p-values from this test are in Supplementary Figure S5 show that the BD-rotamer library results in a statistically meaningful decrease in RMSDs compared to existing methods such as SIDEpro and Rosetta. For all PTM types considered in this work, the PTM BD-rotamer provides excellent performance in recovering the experimental structures; Figure 3B provides examples of repacked PTM-modified structures using all four methods and compared to experimental PDB structures.

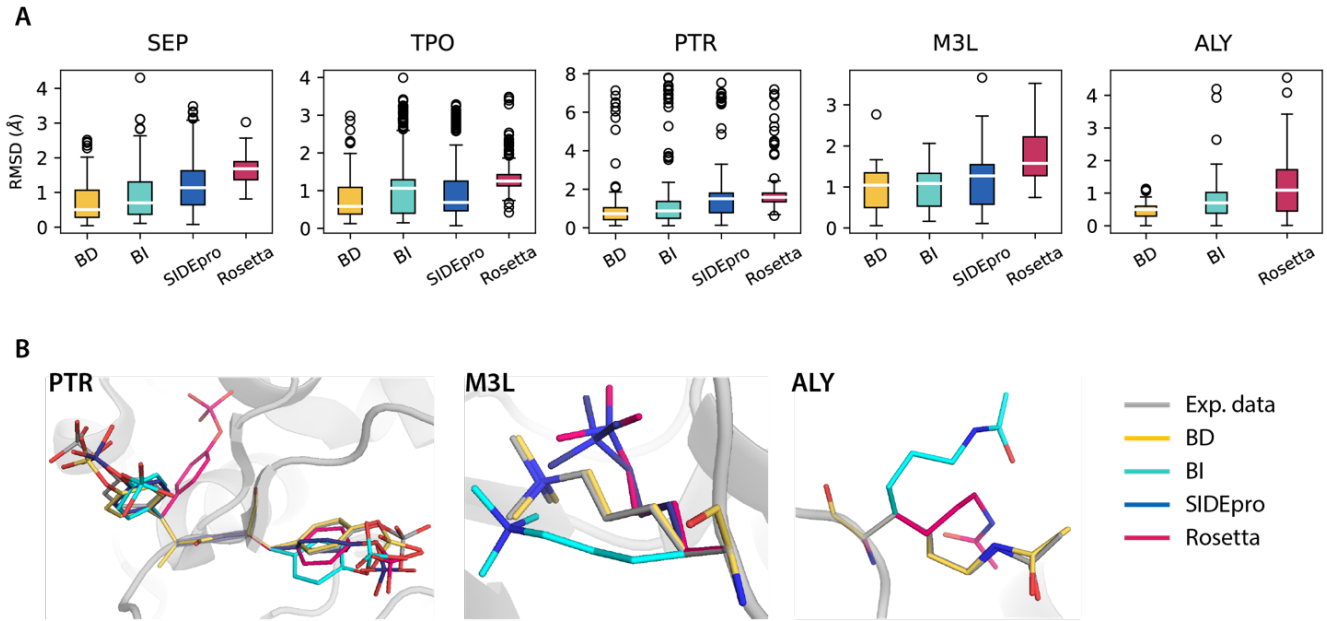

**Figure 3**. *RMSD distributions of repacked PTM-modified residues using different rotamer libraries to experimental structures*. SIDEpro does not support ALY packing. A). Boxplots for the RMSD distributions. Medians are highlighted in white and each box extends from the first quartile (Q1) to the third quartile (Q3). Outliers in circle are defined as points outside of 1.5 times the interquartile range below Q1 or above Q3. B). Repacked PTM-containing structures using different rotamer libraries compared to the experimental PDB structures. Examples are taken from PDB ID 3CLY (PTR), 4EZH (M3L) and 4QUT (ALY).

To demonstrate how our PTM libraries can support ensemble generation for disordered proteins, we considered two cases for which the proteins contain IDRs. The Histone H3 N-terminal IDR within the nucleosome structure (chains C and G, PDB ID 8SIY) have 4 of the 5 types of PTMs investigated here, with methylation at K9, phosphorylation at S10 and T11, and acetylation at K14 (Figure 4A). The ubiquitin recognition factor in ER-associated degradation protein 1 (UFD1) with phosphorylation at Y219 (PDB ID 2YUJ) is shown in Figure 5A. For both cases we compare ensembles generated with and without PTMs using our BD-rotamer and BI-rotamer libraries as well as libraries from SIDEpro and Rosetta, although SIDEpro only contains M3L, SEP and TPO since ALY is not supported by that method[25]. We also compared against results using the original Rosetta scoring function that ignores steric clashes for the Histone H3 nucleosome structure. Supplementary Figure S6 shows that the Ramachandran plots of the sidechain rotamer states with PTMs do not change for IDRs, nor among libraries, and only small changes are observed in secondary structure between rotamer libraries as seen in Supplementary Figure S7. This indicates that the structural changes are concentrated in any differences in side chain packing.

Figures 4C and 5C show that the torsional properties of the PTM IDR ensembles are different to IDR ensembles without sidechain modifications at the modification sites. Furthermore, the sidechain rotamer state changes are reflected differently for the BD-, BI-, SIDEpro, and two Rosetta-rotamer ensembles. To better analyze what are the structural consequences of the IDR ensembles generated with the different PTM rotamer libraries, we constructed 2D contact maps subtracting the values from the ensemble without PTMs (Figures 4D and 5D) to look for increases (red) or decreases (blue) of residue-residue contacts being made. For Histone H3, residues 5-14 show a higher population of contacts with the BD-rotamer library, which we see corresponds to a much denser hydrogen-bonded network on average in this area (Figure 4B). A similar observation is found for UDF1 using the BD-rotamer library, and although it is only a single modification spot at residue 219, the PTM introduces a network of hydrogen bonds (example shown in Figure 5B). Especially for Histone H3, the other rotamer libraries show a smaller set of contacts or even net loss of contacts in this same region. In turn the BD-rotamer generated structures show a diminishment of contacts made by these same PTM residues with other regions of the protein, an effect which is muted in the other libraries. For the BI-rotamer library this is due to over-averaging, whereas for SIDEPro the differences with the unmodified ensembles is because it uses the same $\chi_1$ as the unmodified residues. It is interesting that the Rosetta rotamer-library combined with no clash criteria gives a nearly opposite trend in sidechain packing for residues with PTMs. Given that repacking structures on the folded protein backbone showed greater reliability with the BD-rotamer set, we extrapolate that there is better justification for supporting the structural consequences observed for the Histone H3 and UDF1 IDRs.

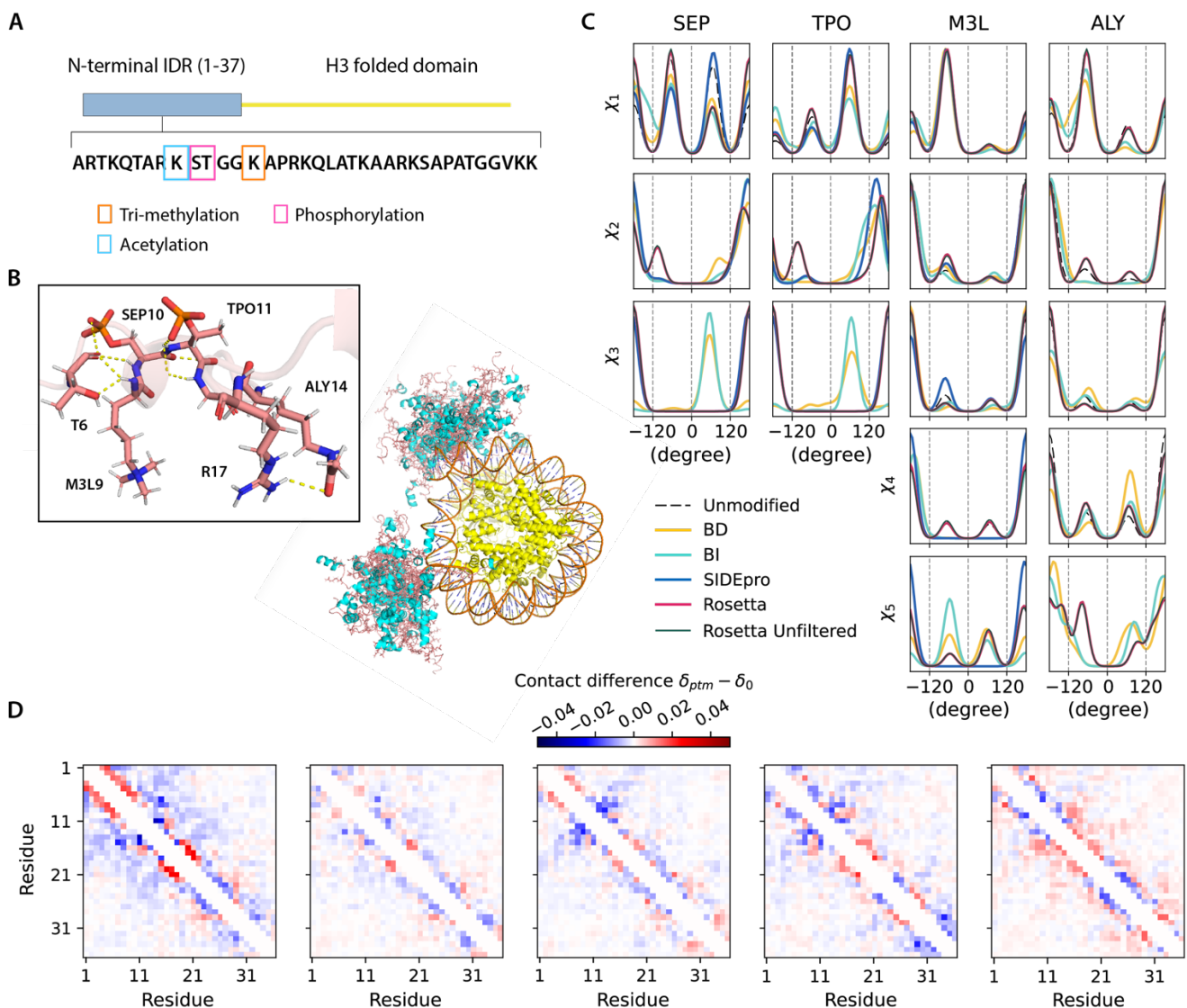

**Figure 4**. *Histone H3 conformers generated with different PTMs libraries.* A). Modifications on the histone H3 N-terminus on chains C/G of the nucleosome. B). Ensembles of 30 all-atom H3-modified nucleosome conformers (folded domains and DNA are taken from PDB ID 8SIY). The N-terminal IDRs on chain C and G are shown in salmon (loop) and cyan (helices), and folded domains (yellow). PTM-modified residues are highlighted with stick representations. C). Comparison of torsion angle probability distributions for PTM-modified residues in the H3 conformers with different libraries. D). Fractional inter-residue contacts ($C_\alpha$-$C_\alpha$ distances within 8 Å) of PTMs containing ensembles ($\delta_{ptm}$) subtracted by the ensemble without modifications ($\delta_0$) for the IDR regions. The maps were calculated with 500 randomly sampled conformers (based on convergence of Rgand averaged over 10 trials (blue-less, red-more). From left to right: BD-rotamer, BI-rotamer, SIDEpro, Rosetta and Rosetta without clash filtering, with contacts averaged from chain C and G.

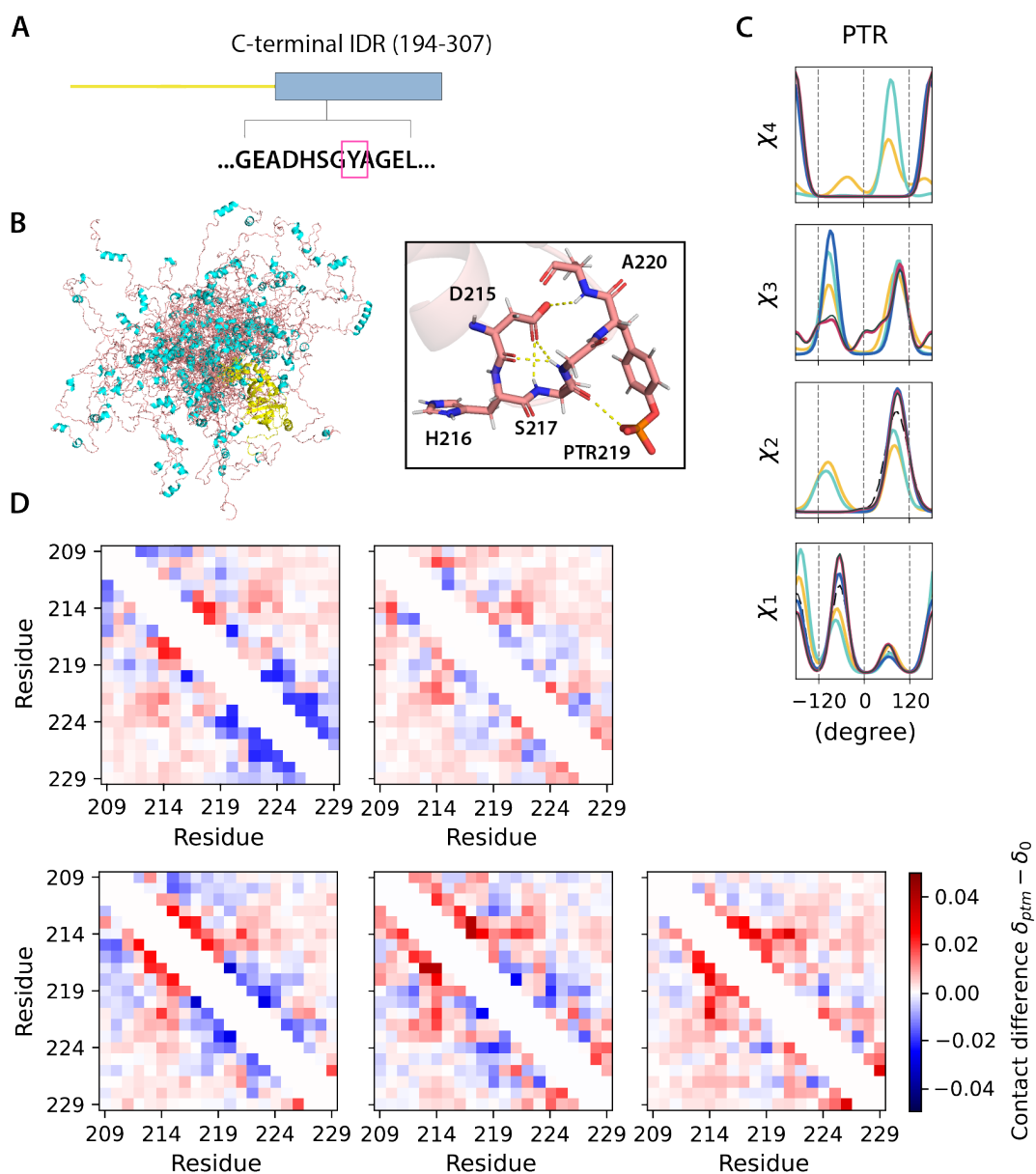

**Figure 5**. *UDF1 conformers generated with different PTM libraries.* A). Modifications on the C-terminal IDR of UDF1 at Y219. B). Cartoon representations of 30 all-atom UDF1 conformers (structure of the folded domain taken from PDB ID 2YUJ model). The C-terminal IDR is shown in salmon red (loop) and cyan (helices), and the folded domains in yellow. C). Comparison of torsion angle probability distributions for PTM in the UDF1 conformers with different libraries. D). Fractional inter-residue contacts ($C_\alpha$-$C_\alpha$ distances within 8 Å) of PTMs containing ensembles ($\delta_{ptm}$) subtracted by the ensemble without modifications ($\delta_0$) for the IDR region around the PTM site. The maps were calculated with ensembles of 500 randomly sampled conformers and averaged over 10 trials. Top: BD-rotamer, BI-rotamer; Bottom (L to R): SIDEpro, Rosetta and Rosetta without clash filtering.

## DISCUSSION AND CONCLUSION

Modeling sidechain conformations for PTMs have important applications in understanding protein conformational switches, including changes in dynamics and disordered protein interactions, leading to functional consequences modulated by PTMs. Although publicly available high-resolution structures including residues modified by many of the known PTMs are still limited, there is enough structural data for residues modified by phosphorylation, methylation, and acetylation. Thus, we have developed BD-rotamer and BI-rotamer libraries for PTM-modified residues, using structural data curated and cleaned from the recent PTM-SD update.

We evaluate the constructed libraries in comparison to SIDEpro and Rosetta in the context of generating conformers for folded domains and for proteins with disordered regions. The BD-rotamer libraries outperform the BI-rotamer libraries as well as SIDEpro and Rosetta in retrieving the experimental structures for PTMs on the folded proteins, as evidence of its ability to capture the correlation more accurately between backbone and sidechain, and within sidechain dihedral conformations. For phosphorylated residues specifically, the ability to predict the sidechain rotamer cooperatively is crucial to modeling local hydrogen bonding interactions, thus allowing one to better delineate the structural features of single conformer and ensembles upon chemical modifications. We also show that the constructed PTM-modified residue rotamer libraries can be used along with MCSCE and IDPConformerGenerator to produce all-atom conformers for disordered proteins with PTMs. If experimental data such as nuclear magnetic resonance, small-angle X-ray scattering, and fluorescence resonance energy transfer are available, a reweighting protocol using X-EISD[46] or evolving underlying ensembles to agree with experimental data using DynamICE[42] can be applied to these all-atom sidechain-modified conformers to generate more realistic ensemble representations to investigate PTM-regulated activities and interactions.

## ACKNOWLEDGEMENTS

J.D.F.-K. and T.H.-G. acknowledge funding from the National Institutes of Health (2R01GM127627-05). J.D.F.-K. also acknowledges support from the Natural Sciences and Engineering Research Council of Canada (NSERC, 2016-06718) and from the Canada Research Chairs Program.

## DATA AND CODE AVAILABILITY

The codes for dihedral angle computations and library creation are available at https://github.com/THGLab/ptm_sc.git. They support backbone dependent and independent rotamer analysis and library construction and can be extended to other PTMs. We also note that the library is part of the MCSCE program seamlessly integrated into the IDPConformerGenerator platform, where users can generate conformers with disordered regions with PTMs in one command line.

## REFERENCES

1   Ramazi, S. & Zahiri, J. Post-translational modifications in proteins: resources, tools and prediction methods. *Database* **2021**, baab012, doi:10.1093/database/baab012 (2021).
2   Huang, K.-Y. *et al.* dbPTM in 2019: exploring disease association and cross-talk of post-translational modifications. *Nucleic Acids Research* **47**, D298-D308, doi:10.1093/nar/gky1074 (2019).
3   Jumper, J. *et al.* Highly accurate protein structure prediction with AlphaFold. *Nature* **596**, 583-589, doi:10.1038/s41586-021-03819-2 (2021).
4   Baek, M. *et al.* Accurate prediction of protein structures and interactions using a three-track neural network. *Science* **373**, 871-876, doi:10.1126/science.abj8754 (2021).
5   Lane, T. J. Protein structure prediction has reached the single-structure frontier. *Nature Methods* **20**, 170-173, doi:10.1038/s41592-022-01760-4 (2023).

6       Wolff, A. M. *et al.* Mapping protein dynamics at high spatial resolution with temperature-jump X-ray crystallography. *Nature Chemistry* **15**, 1549-1558, doi:10.1038/s41557-023-01329-4 (2023).
7       Wayment-Steele, H. K. *et al.* Predicting multiple conformations via sequence clustering and AlphaFold2. *Nature* **625**, 832-839, doi:10.1038/s41586-023-06832-9 (2024).
8       Uversky, V. N., Oldfield, C. J. & Dunker, A. K. Intrinsically Disordered Proteins in Human Diseases: Introducing the D2 Concept. *Annual Review of Biophysics* **37**, 215-246 (2008).
9       Wright, P. E. & Dyson, H. J. Intrinsically disordered proteins in cellular signalling and regulation. *Nat Rev Mol Cell Biol* **16**, 18-29 (2015).
10      Bhowmick, A. *et al.* Finding Our Way in the Dark Proteome. *J Am Chem Soc* **138**, 9730-9742, doi:10.1021/jacs.6b06543 (2016).
11      Lazar, T. *et al.* PED in 2021: a major update of the protein ensemble database for intrinsically disordered proteins. *Nucleic Acids Research* **49**, D404-D411, doi:10.1093/nar/gkaa1021 (2021).
12      Ghafouri, H. *et al.* PED in 2024: improving the community deposition of structural ensembles for intrinsically disordered proteins. *Nucleic Acids Research* **52**, D536-D544, doi:10.1093/nar/gkad947 (2024).
13      Fraser, J. S. *et al.* Accessing protein conformational ensembles using room-temperature X-ray crystallography. *Proceedings of the National Academy of Sciences of the United States of America* **108**, 16247-16252, doi:10.1073/pnas.1111325108 (2011).
14      Moorman, V. R., Valentine, K. G. & Wand, A. J. The dynamical response of hen egg white lysozyme to the binding of a carbohydrate ligand. *Protein science : a publication of the Protein Society* **21**, 1066-1073, doi:10.1002/pro.2092 (2012).
15      Fenwick, R. B., van den Bedem, H., Fraser, J. S. & Wright, P. E. Integrated description of protein dynamics from room-temperature X-ray crystallography and NMR. *Proceedings of the National Academy of Sciences of the United States of America* **111**, E445-454, doi:10.1073/pnas.1323440111 (2014).
16      Welborn, V. V. & Head-Gordon, T. Fluctuations of Electric Fields in the Active Site of the Enzyme Ketosteroid Isomerase. *Journal of the American Chemical Society* **141**, 12487-12492, doi:10.1021/jacs.9b05323 (2019).
17      Richard, J. P. Protein Flexibility and Stiffness Enable Efficient Enzymatic Catalysis. *Journal of the American Chemical Society* **141**, 3320-3331, doi:10.1021/jacs.8b10836 (2019).
18      Bhowmick, A. & Head-Gordon, T. A Monte Carlo method for generating side chain structural ensembles. *Structure* **23**, 44-55, doi:http://dx.doi.org/10.1016/j.str.2014.10.011 (2015).
19      Dicks, L. & Wales, D. J. Exploiting Sequence-Dependent Rotamer Information in Global Optimization of Proteins. *The Journal of Physical Chemistry B* **126**, 8381-8390, doi:10.1021/acs.jpcb.2c04647 (2022).
20      Jumper, J. M., Faruk, N. F., Freed, K. F. & Sosnick, T. R. Accurate calculation of side chain packing and free energy with applications to protein molecular dynamics. *PLOS Computational Biology* **14**, e1006342, doi:10.1371/journal.pcbi.1006342 (2018).
21      Liang, S., Zheng, D., Zhang, C. & Standley, D. M. Fast and accurate prediction of protein side-chain conformations. *Bioinformatics* **27**, 2913-2914 (2011).
22      Ollikainen, N., de Jong, R. M. & Kortemme, T. Coupling Protein Side-Chain and Backbone Flexibility Improves the Re-design of Protein-Ligand Specificity. *PLOS Computational Biology* **11**, e1004335, doi:10.1371/journal.pcbi.1004335 (2015).
23      Huang, X., Pearce, R. & Zhang, Y. FASPR: an open-source tool for fast and accurate protein side-chain packing. *Bioinformatics* **36**, 3758-3765, doi:10.1093/bioinformatics/btaa234 (2020).
24      McPartlon, M. & Xu, J. An end-to-end deep learning method for protein side-chain packing and inverse folding. *Proceedings of the National Academy of Sciences* **120**, e2216438120, doi:10.1073/pnas.2216438120 (2023).
25      Nagata, K., Randall, A. & Baldi, P. SIDEpro: a novel machine learning approach for the fast and accurate prediction of side-chain conformations. *Proteins* **80**, 142-153, doi:10.1002/prot.23170 (2012).
26      Misiura, M., Shroff, R., Thyer, R. & Kolomeisky, A. B. DLPacker: Deep learning for prediction of amino acid side chain conformations in proteins. *Proteins: Structure, Function, and Bioinformatics* **90**, 1278-1290, doi:https://doi.org/10.1002/prot.26311 (2022).
27      Petrovskiy, D. V. *et al.* Modeling Side Chains in the Three-Dimensional Structure of Proteins for Post-Translational Modifications. *International Journal of Molecular Sciences* **24**, 13431 (2023).
28      Leaver-Fay, A. *et al.* ROSETTA3: an object-oriented software suite for the simulation and design of macromolecules. *Methods Enzymol* **487**, 545-574, doi:10.1016/b978-0-12-381270-4.00019-6 (2011).

29. Alford, R. F. *et al.* The Rosetta All-Atom Energy Function for Macromolecular Modeling and Design. *Journal of Chemical Theory and Computation* **13**, 3031-3048, doi:10.1021/acs.jctc.7b00125 (2017).
30. Schymkowitz, J. *et al.* The FoldX web server: an online force field. *Nucleic Acids Research* **33**, W382-W388, doi:10.1093/nar/gki387 (2005).
31. Renfrew, P. D., Craven, T. W., Butterfoss, G. L., Kirshenbaum, K. & Bonneau, R. A rotamer library to enable modeling and design of peptoid foldamers. *J Am Chem Soc* **136**, 8772-8782, doi:10.1021/ja503776z (2014).
32. Craveur, P., Rebehmed, J. & de Brevern, A. G. PTM-SD: a database of structurally resolved and annotated posttranslational modifications in proteins. *Database* **2014**, bau041, doi:10.1093/database/bau041 (2014).
33. Berman, H. M. *et al.* The Protein Data Bank. *Nucleic Acids Research* **28**, 235-242, doi:10.1093/nar/28.1.235 (2000).
34. Altschul, S. F. *et al.* Gapped BLAST and PSI-BLAST: a new generation of protein database search programs. *Nucleic Acids Res* **25**, 3389-3402, doi:10.1093/nar/25.17.3389 (1997).
35. Shapovalov, Maxim V. & Dunbrack, Roland L., Jr. A Smoothed Backbone-Dependent Rotamer Library for Proteins Derived from Adaptive Kernel Density Estimates and Regressions. *Structure* **19**, 844-858, doi:10.1016/j.str.2011.03.019 (2011).
36. Mardia, K. V. & Zemroch, P. J. The Von Mises Distribution Function. *Journal of the Royal Statistical Society Series C: Applied Statistics* **24**, 268-272, doi:10.2307/2346578 (1975).
37. Nadaraya, E. A. On Estimating Regression. *Theory of Probability & Its Applications* **9**, 141-142, doi:10.1137/1109020 (1964).
38. Watson, G. S. Smooth Regression Analysis. *Sankhyā: The Indian Journal of Statistics, Series A (1961-2002)* **26**, 359-372 (1964).
39. Liu, Z. H. *et al.* Local Disordered Region Sampling (LDRS) for ensemble modeling of proteins with experimentally undetermined or low confidence prediction segments. *Bioinformatics* **39**, btad739, doi:10.1093/bioinformatics/btad739 (2023).
40. Teixeira, J. M. C. *et al.* IDPConformerGenerator: A Flexible Software Suite for Sampling the Conformational Space of Disordered Protein States. *The Journal of Physical Chemistry A* **126**, 5985-6003, doi:10.1021/acs.jpca.2c03726 (2022).
41. Dunbrack Jr, R. L. & Cohen, F. E. Bayesian statistical analysis of protein side-chain rotamer preferences. *Protein Science* **6**, 1661-1681, doi:https://doi.org/10.1002/pro.5560060807 (1997).
42. Zhang, O. *et al.* Learning to evolve structural ensembles of unfolded and disordered proteins using experimental solution data. *J Chem Phys* **158**, doi:10.1063/5.0141474 (2023).

# Supplementary Information
# A Curated Rotamer Library for Common Post-Translational Modifications of Proteins

Oufan Zhang[1], Shubhankar A. Naik[2], Zi Hao Liu[5,6], Julie Forman-Kay[5,6],
Teresa Head-Gordon[1-4*]

[1]Kenneth S. Pitzer Center for Theoretical Chemistry, [2]Department of Chemistry,
[3]Department of Bioengineering, and [4]Department of Chemical and Biomolecular Engineering,
University of California, Berkeley, Berkeley, California 94720, USA

[5]Molecular Medicine Program, Hospital for Sick Children, Toronto, Ontario M5G 0A4, Canada,
[6]Department of Biochemistry, University of Toronto, Toronto, Ontario M5S 1A8, Canada

## Supplementary Tables

**Table S1**. *Number of PDB files for PTM-modified proteins and their resolution ranges.* PDB files were found for phosphorylated serine (SEP), phosphorylated threonine (TPO), phosphorylated tyrosine (PTR), methylated arginine (AGM), mono-methylated lysine (MLZ), di-methylated lysine (MLY), tri-methylated lysine (M3L), oxidized methionine (OMT), and acetylated lysine (ALY). All PDB identifiers are provided in Table S2. Based on a trade-off between data quality and data abundance, we have only developed rotamer libraries for SEP, TPO, PTR, M3L, and ALY. See Methods for more details.

| Refinement Resolution (Å) | SEP | TPO | PTR | AGM | MLZ | MLY | M3L | OMT | ALY |
|---|---|---|---|---|---|---|---|---|---|
| **1.0 - 1.5** | 6 | 2 | 4 | 6 | 2 | 1 | 2 | —— | 1 |
| **1.5 - 2.0** | 68 | 57 | 37 | 4 | 11 | 5 | 33 | —— | 19 |
| **2.0 - 2.5** | 115 | 132 | 84 | —— | 3 | 5 | 20 | —— | 10 |
| **2.5 - 3.0** | 46 | 56 | 43 | —— | 2 | —— | 6 | 1 | 9 |
| **3.0 - 3.5** | 8 | 25 | 8 | —— | —— | —— | —— | —— | —— |
| **3.5 - 4.0** | 6 | 8 | —— | —— | —— | —— | —— | —— | —— |
| **4.0 - 4.5** | 1 | 2 | —— | —— | —— | —— | —— | —— | —— |
| **> 4.5** | —— | —— | —— | —— | —— | 8 | —— | —— | —— |
| **TOTAL** | 250 | 282 | 176 | 10 | 18 | 19 | 61 | 1 | 39 |

**Table S2**. *PDB ID codes for proteins containing PTM-modified residues with a 3.5 Å resolution cutoff.*

| PTM | PDB ID |
|---|---|
| SEP | 1ATP;1L3R;1APM;1FMO;1CMK;1BKX;1BX6;1JBP;1J3H;1LWO;1KHX;1OVA;1KKM;1I7W;1JLU;1SVG;1RDQ;1Q8U;1Q8W;1STC;1SMH;1REK;1SZM;1SVH;1SYK;1U7F;1YDR;1U7V;1U5R;1XJD;1UHG;1UJK;1VEB;2BVA;1ZEF;1ZEB;1Z8D;2BIK;1YHS;1YDS;1YDT;2C1A;2C1J;2CPK;2C1N;2C1B;2ERZ;2CDZ;2GNG;2GNL;2FEP;2GNF;2GLQ;2GNJ;2GFC;2GPA;2GNH;2GNI;2JDV;2OEN;2OBJ;2NZV;2JED;2JFL;2NRU;2JDT;2NZU;2J0I;2QCS;2OIB;2QLL;2QVS;2OIC;2QUR;2Q0N;2OID;2UVY;2UVX;3AGM;2XIX;2VNY;2XIY;2UVZ;3AMA;2VNW;2XCK;2VO0;2XIZ;3COJ;3AMB;3DNE;3BGZ;3BGM;3BWJ;3BGP;3DND;3DDW;3BGQ;3G2V;3FQR;3FHI;3F2A;3E8E;3EXH;3FQX;3FJQ;3FQU;3G2T;3HUF;3H9O;3IDC;3HRF;3HRC;3IQJ;3IDB;3IFQ;3IOP;3L9N;3NKX;3NAY;3KKV;3NUN;3IQU;3IQV;3L9M;3L9L;3L6F;3O8I;3OQM;3O7L;3OQN;3ORZ;3NUS;3OTU;3ORX;3NUY;3NUU;3QAM;3OVV;3P0M;3OW3;3PVB;3QCS;3POO;3QAL;3Q4A;3OXT;3QD3;3SQD;3T7K;3T9I;3RCJ;3RWP;3QCX;3RWQ;3SZM;3TPE;3UIM;3TMP;3UAL;3UZD;3UEO;3ULZ;3TPV;3U3Z;3UBW;4BZN;3ZKF;3VQH;3X2W;3X2V;4C0O;4BZO;3V7D;4FIG;4FIE;4FIF;4FII;4FIJ;4DAU;4EUU;4FIH;4DG0;4DFY;4IAI;4IAK;4HPT;4IAZ;4IAY;4IAD;4HPU;4IAC;4IAF;4IGK;4JAX;4JDH;4JDJ;4KIK;4IHL;4JDK;4N6Z;4N6Y;4JDI;4O0V;4OTI;4OTG;4O46;4OTH;4O0Y;4PSI;4NTS;4O0X;4NU1;4QPM;4RRV;4RQV;4UJB;4RQK;4WB8;4Q9Z;4UJ1;4RMZ;4XS2;4YP8;4YO6;4ZTM;5ACK;5K7I;5K7G;4ZTL;4ZTN;5K75;5T1S;5T1T;5UPL;5UZK;5K72;5K76;5VED;5UPK;5TOS;5VEE;5VEF |
| TPO | 1H1Q;1L3R;1JSU;1JLU;1JST;1FMO;1H1S;1H1R;1FOT;1JBP;1QMZ;1O6K;1Q8W;1O6L;1PKD;1Q8U;1RE8;1RDQ;1P5E;1SZM;1W98;1SVG;1SVH;1U9I;1SMH;1REJ;1STC;1REK;1SYK;2A1A;1ZRZ;2B2T;1YDS;1YDT;1YRP;1YDR;1XJD;2CPK;2ERZ;2CCI;2ERK;2CCH;2C1B;2C6T;2C1A;2GBL;2G9X;2GNG;2GNF;2GFC;2JDV;2JFL;2GNI;2GNJ;2JDT;2IW6;2JDS;2GNL;2JDR;2GNH;2QCS;2JGZ;2NRU;2OIC;2NRY;2JFM;2OIB;2Q8Y;2OID;2O8Y;2QVS;2UZD;2VNW;2QUR;2UVX;2VNY;2VAG;2UVZ;2UVY;3A8X;2WMB;3A7J;3AGM;3A7H;3A8W;3AL3;2WMA;3A7G;3A7I;3AMB;3BLR;3BHT;3BLH;3BHV;3BHU;3AMA;3BLQ;3COM;3CKX;3D0E;3BZI;3CQW;3C4W;3BWJ;3CQU;3FHI;3DNE;3DVL;3DDP;3DND;3DOG;3E87;3E5A;3E8E;3DDQ;3FXZ;3K0E;3K0C;3K0A;3IDC;3FJQ;3IDB;3MI9;3K2L;3OJY;3O7L;3K09;3L9N;3L9M;3L9L;3MVJ;3POO;3PVB;3OW3;3P0M;3QAL;3OOG;3OXT;3QHW;3QAM;3OVV;3QKK;3UIM;3TN8;3TNH;3TXO;3TNI;3UIG;3TNW;3UEO;4BCF;4A0N;4BCG;4BCH;3UOT;3ULZ;4A0J;3UNN;3X2W;3X2V;4BCI;4C4F;4BCN;4BCK;4BCQ;4BCM;4BCO;4C4G;4BCJ;4BCP;4CRS;4CFU;4CXA;4CFW;4CEG;4CFV;4CFE;4CFH;4CFM;4CFN;4EOJ;4DUG;4DC2;4EOK;4DAW;4ELJ;4EOI;4EKL;4EOL;4EOR;4EOP;4EOO;4EON;4EOS;4EQC;4EOQ;4EWQ;4EOM;4IAZ;4IAC;4GLR;4IAK;4I3Z;4IAF;4IAD;4HPU;4IAI;4IAY;4OR5;4OTG;4IMY;4IZA;4NTS;4IJM;4NST;4OGR;4OTD;4QBS;4QMM;4QML;4QFG;4QMN;4QMO;4QMS;4QMP;4QFR;4QMQ;4QMU;4QMW;4QMX;4QMT;4RMZ;4RA4;4QMV;4QNA;4QMY;4QMZ;4ZTN;4XS2;4ZTM;4YO6;4YP8;4WB8;4UJB;4ZHX;4ZTL;4U8Z;5K72;5K7G;5K76;5K7I;5K75;5T1T;5T1S;5L2W;5KQ5;5UZK;6C9G;6ATH;5T5T;5V60;6BDL;5V61;6BG2;1APM;8H6P;1CMK;1ATP;1BX6;1BKX;8H6T;6C9J |
| PTR | 3CLY;3BUX;3C7Q;3EB0;3BYO;3DQW;3BUW;3BYM;3CD3;3KCK;3EYH;3F7Z;3FUP;3F5P;3KVW;3KMM;3KRR;3LCK;3LXN;3KXZ;3MXY;3LPB;3MAZ;3LXP;3RVG;3SRV;3SAY;3PFV;3R7O;3NNX;3TJC;3OP0;3TJD;3VRO;3ZEW;3VRZ;3VRY;3VRP;3U3Z;3VRR;3ZEP;4CKJ;3ZUV;4AFJ;3ZMM;4CKI;4A4C;3ZNI;4AZF;3ZRK;4AQC;4E4L;4E5W;4E4M;4DIT;4EHZ;4E4N;4E6Q;4D1S;4E6D;4EI4;4F09;4GFM;4FK6;4GVC;4GFU;4HGE;4GPL;4I5C;4F08;4IVA;4IFC;4IZA;4IIR;4IAN;4IJP;4IVC;4IVD;4IWD;4IVB;4J1R;4K11;4NCT;4JMG;4LUE;4JI9;4JIA;4K6Z;4JMH;5A3X;4R3P;4R3R;4RXZ;4S0G;4Z16;4V0G;4QUM;5TQ6;5A54;5A4T;5A4L;5AMN;5CZI;5TQ3;5A4Q;5AIK;5AEP;5V60;5TQ7;6C7Y;5V61;5WEV;6CZ2;5USY;1FMK;1F1W;1AYA;1FBV;1AYB;1AYC;1I3Z;1PKG;1NZV;1GAG;1NZL;1LCJ;1O9U;1KSW;1LCK;1SPS;1QPJ;1SHA;1RQQ;1QPE;1QCF;1TRN;1UUR;1ZFP;1UUS;1YVJ;1YVH;2B7A;1YWN;2AUH;2B4S;2C0O;2GQG;2C0I;2H8H;2C0T;2ERK;2CBL;2HDX;2HCK;2IVV;2IVT;2HMH;2IUH;2IVU;2OFU;2OH4;2J0L;2VIF;2PTK;2Q8Y;2SRC;2QOQ;2X2L;2QON;2PVF;3BUN;2ZYB;3BUO;3ANR;2X2M;2Z8C |
| M3L | 1CSX;1CIG;1CRH;1CRI;1CSW;1CSU;1CRG;1CIH;1CSV;1CRJ;1IRV;1PDQ;1KNE;1IRW;1CTY;1KYO;1CTZ;1PRW;2B2W;1RAP;2B2T;2B2U;1RAQ;1YCC;2X4X;3AVR;2X4W;2YCC;2X4Y;3G7L;3LQJ;3K |
|  | QI;3CX5;3N9P;3U5O;3N9Q;3U5N;3N9N;3N9O;3QBY;4EZH;4IUR;3U5P;4N4I;4N4H;4MZH;4NW2;4LXL;4MZF;4MZG;6C4W;4V2V;6AT0;6ASZ;4V2W;1CHI;1CHH;1CIE;1CCR;1CHJ;1CIF |
| ALY | 4LLB;4DNC;3U5P;3UVY;3UVX;4N4F;4QUT;3UVW;4N3W;4QUU;1SZC;4QYL;4QYD;2B5G;2H4F;1SZD;2GIV;2H2D;2E3K;2C1J;2QQF;2QQG;2OU2;2OD9;2OD7;2H4H;2I2Z;2YBG;2Y0M;3JVK;3GLR;3QZT;3JR3;3QZS;3U5N;3TOA;3U5O;3TOB;3QZV |

Table S3. *Side-chain angle definitions for PTM-modified residues.*

|  | $\chi_1$ | $\chi_2$ | $\chi_3$ | $\chi_4$ | $\chi_5$ |
|---|---|---|---|---|---|
| SEP | N-CA-CB-OG | CA-CB-OG-P | CB-OG-P-O1P | ———— | ———— |
| TPO | N-CA-CB-OG1 | CA-CB-OG1-P | CB-OG1-P-O1P | ———— | ———— |
| PTR | N-CA-CB-CG | CA-CB-CG-CD1 | CE1-CZ-OH-P | CZ-OH-P-O1P | ———— |
| M3L | N-CA-CB-CG | CA-CB-CG-CD | CB-CG-CD-CE | CG-CD-CE-NZ | CD-CE-NZ-CM1 |
| ALY | N-CA-CB-CG | CA-CB-CG-CD | CB-CG-CD-CE | CG-CD-CE-NZ | CD-CE-NZ-CH |

Table S4. *Number of successful conformer builds with different rotamer libraries for each case of a protein containing intrinsically disordered protein regions.*

|  | H4K9me3 | UFD1 |
|---|---|---|
| **Unmodified** | 1348 | 2009 |
| **Bb. dp.** | 854 | 1963 |
| **Bb. indp.** | 1266 | 2227 |
| **SIDEpro** | 1192 | 1998 |
| **Rosetta (clash filtered)** | 1359 | 1256 |

Table S5. *RMSDs of repacked PTM-modified residues using different rotamer libraries to experimental structures.* The numbers of PDB files evaluated are shown in parenthesis. SIDEpro does not support ALY packing.

| PTM Library | ALY (31) | M3L (57) | SEP (226) | TPO (256) | PTR (156) |
|---|---|---|---|---|---|
| **BD-rotamers** | 0.49 ± 0.28 | 0.96 ± 0.67 | 0.67 ± 0.53 | 0.76 ± 0.40 | 1.06 ± 1.29 |
| **BI-rotamers** | 0.96 ± 0.99 | 1.01 ± 0.50 | 0.92 ± 0.69 | 1.13 ± 0.88 | 1.56 ± 1.94 |
| **SIDEpro** |  | 1.20 ± 0.73 | 1.24 ± 0.77 | 1.06 ± 0.90 | 1.78 ± 1.68 |
| **Rosetta** | 1.40 ± 1.19 | 1.77 ± 0.67 | 1.66 ± 0.35 | 1.36 ± 0.47 | 1.86 ± 1.14 |

# Supplementary Figures

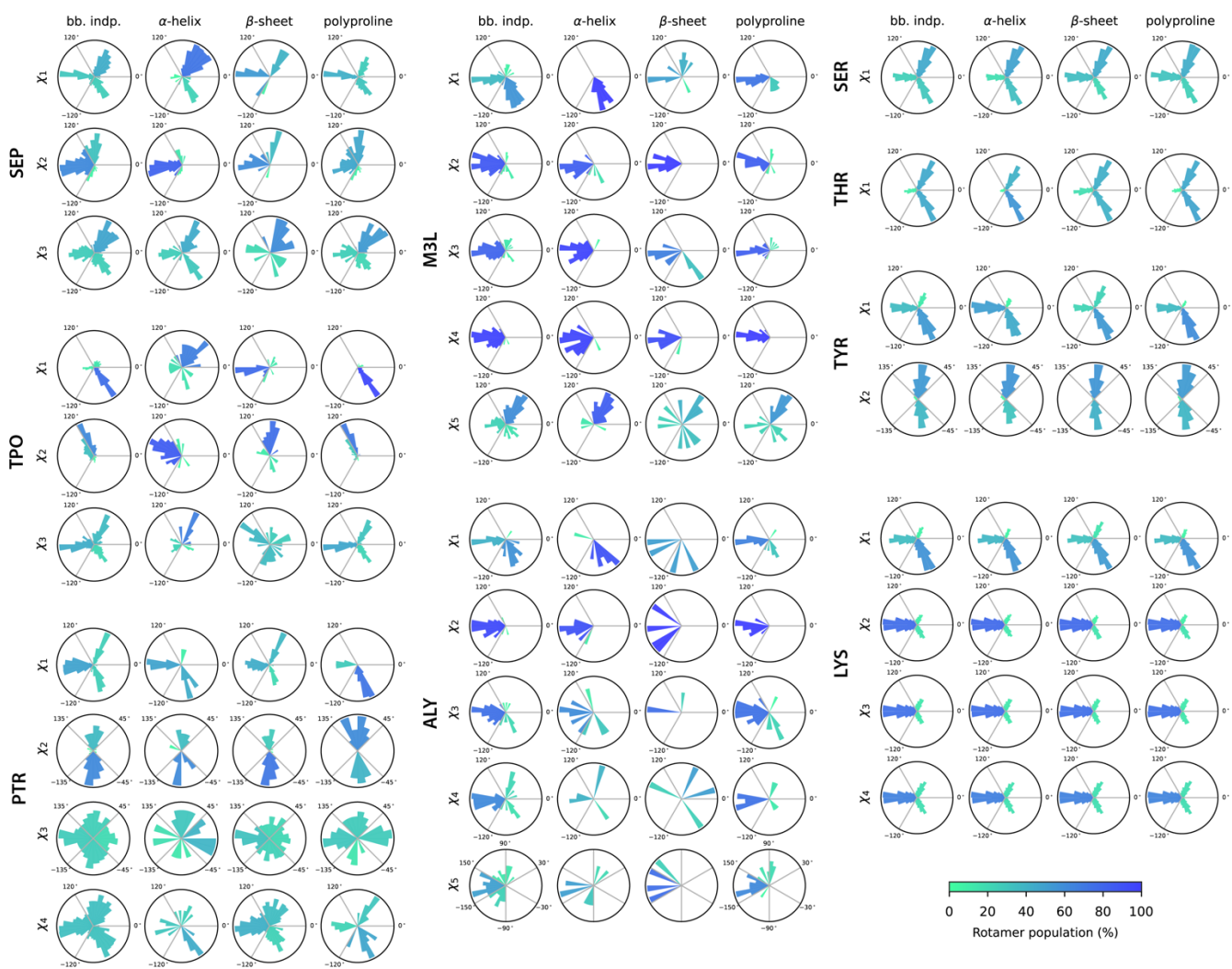

**Figure S1**. *Polar torsion angle histograms and rotameric bin populations for PTM-modified residues in major ϕ/ψ regions for folded proteins.* ϕ/ψ regions are defined as α-helix: -105° ≥ ϕ > -45°, -60° ≥ ψ > 30°; β-sheet: ϕ < -105°, ψ ≥ 90°; polyproline: -105° ≥ ϕ > -45°, ψ ≥ 105°. The backbone independent (bb. indp.) columns are based on all ϕ/ψ regions. Histograms are normalized such that the total area sums to 1. The distributions for canonical amino acids are calculated from SidechainNet's CASP12 dataset with 30% thining[1].

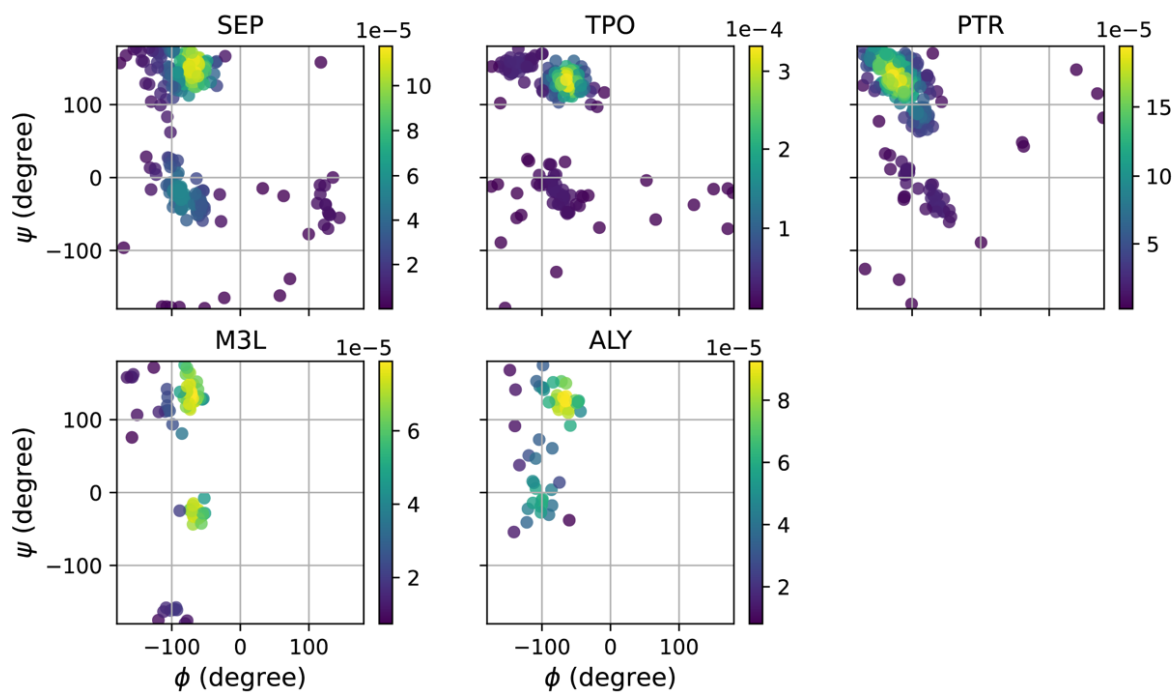

**Figure S2**. *Ramachandran plots for the curated PTM-modified residues.*

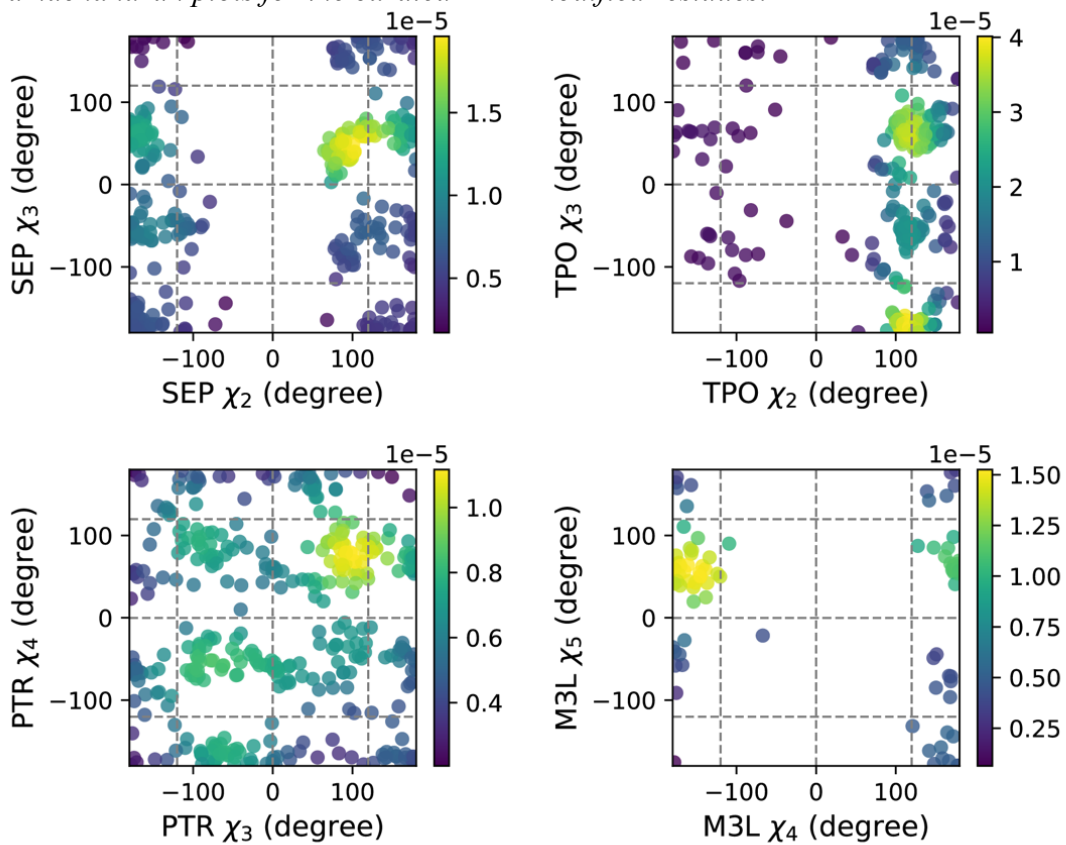

**Figure S3**. *Terminal torsion angle distributions for the curated PTM-modified residues.* Rotameric bin definitions for tetrameric (sp$^3$ bond hybridization) conformations are marked in gray dash line.

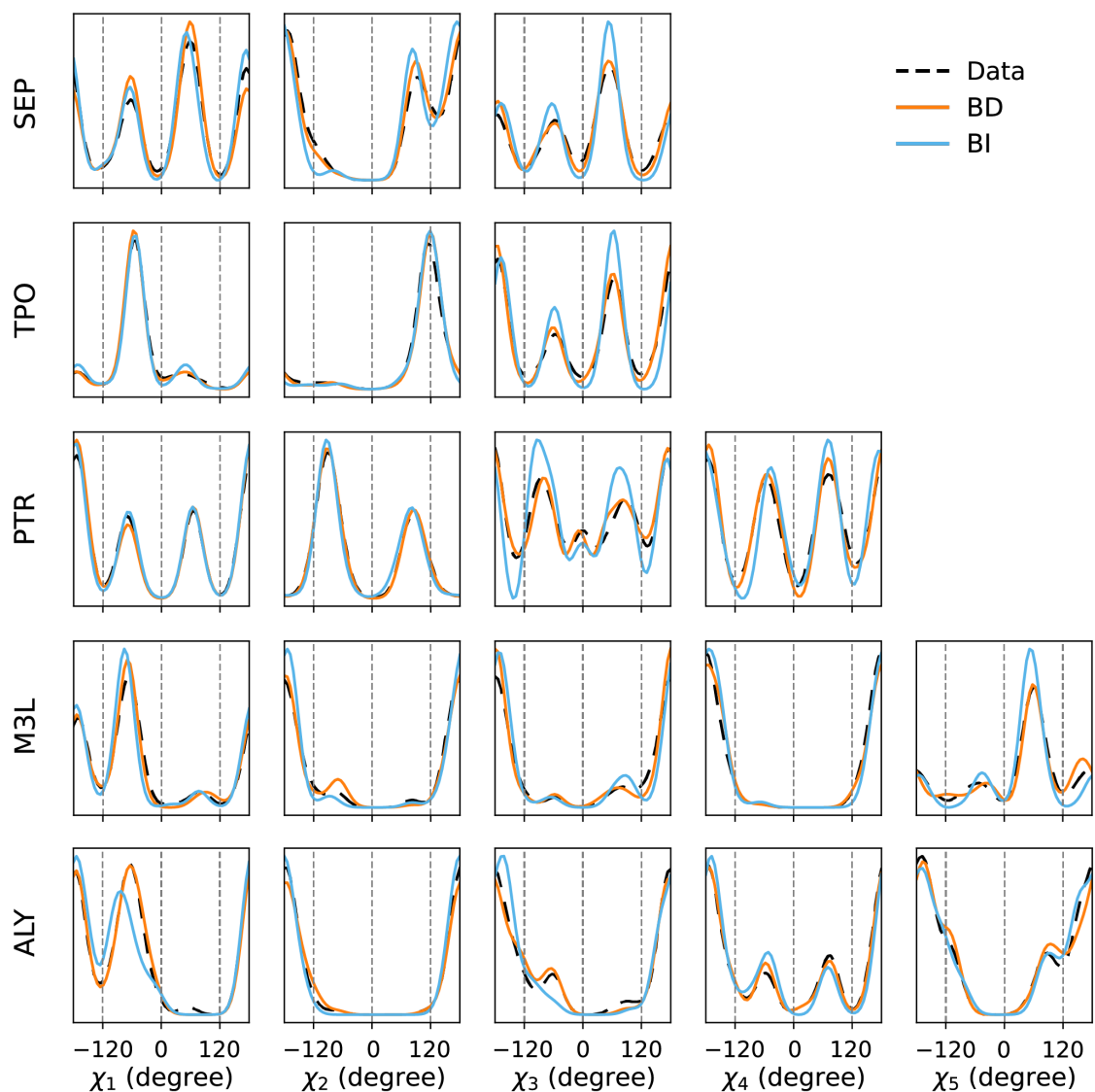

**Figure S4**. *Comparison of rotamer probability distributions for PTM-modified residues from curated PDB data and constructed libraries.* All angles are sampled using the same set of ϕ/ψ, and probability densities are estimated with von Mises kernel.

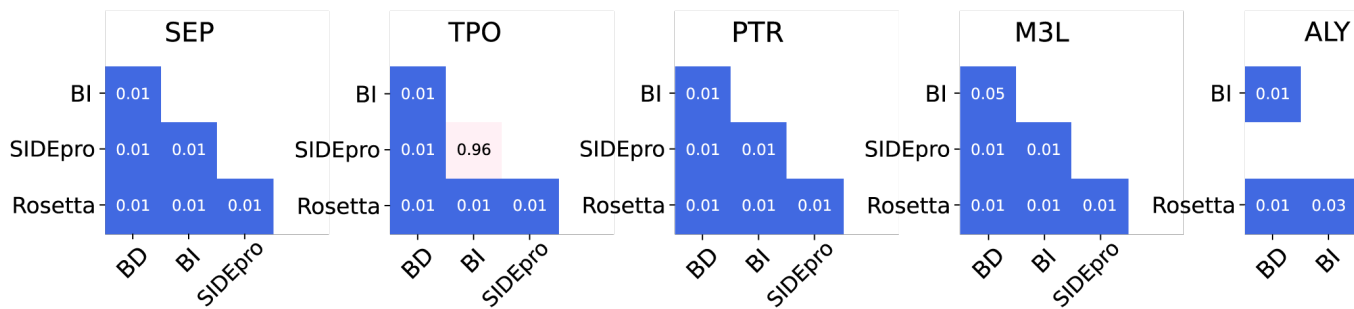

**Figure S5**. *Wilcoxon signed-rank tests for the RMSD distributions of repacked PTM-modified residues to experimental structures.* The one-tailed p-values smaller than 0.05 indicate that the of median of the pairwise difference between the bottom and left distribution is smaller than 0 with a 95% confidence level (highlighted in blue).

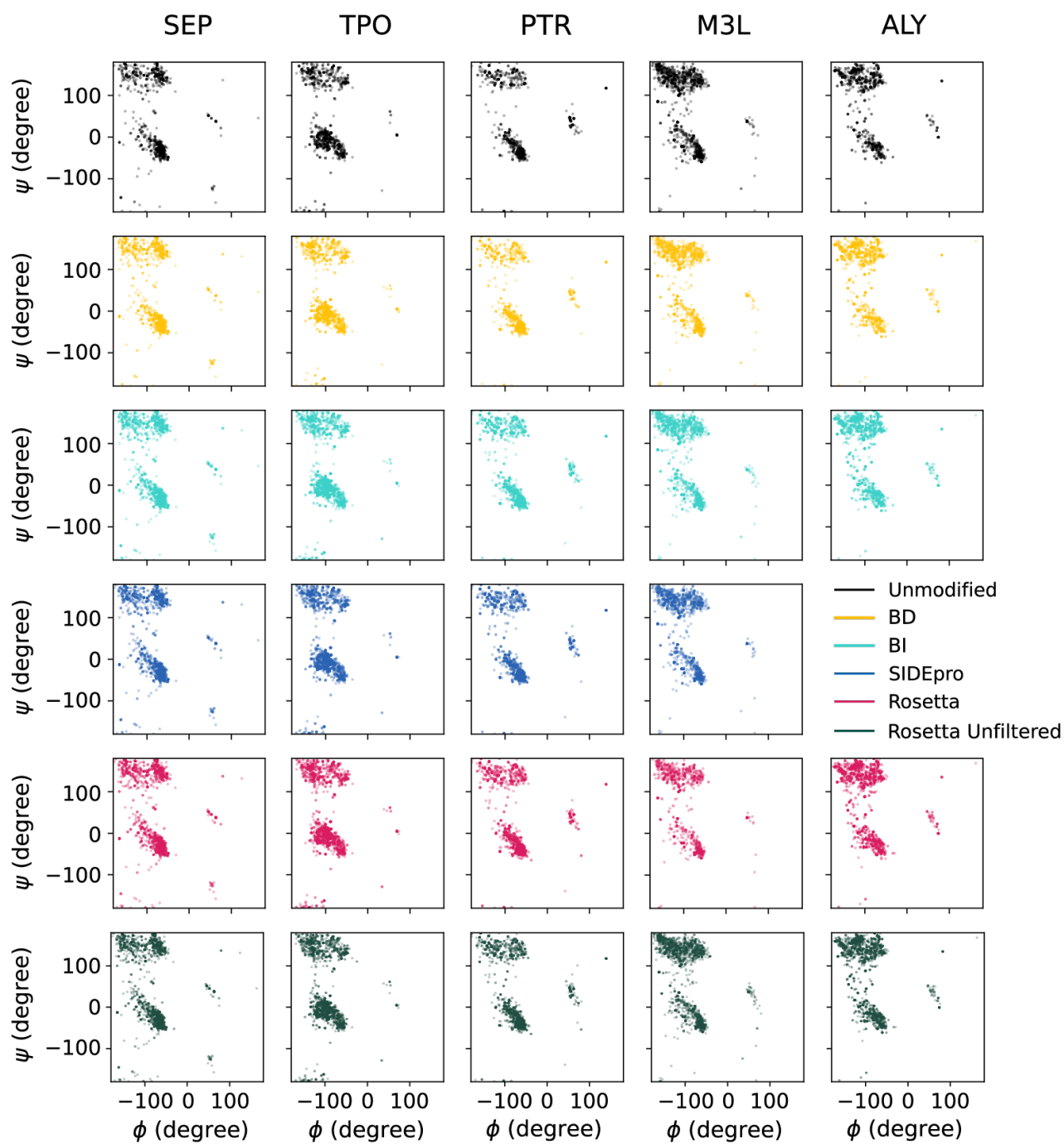

**Figure S6**. *Ramachandran plots of PTM-modified residues in the generated IDR-containing Histone H3 and UDF1 conformers with IDRs using different rotamer libraries.*

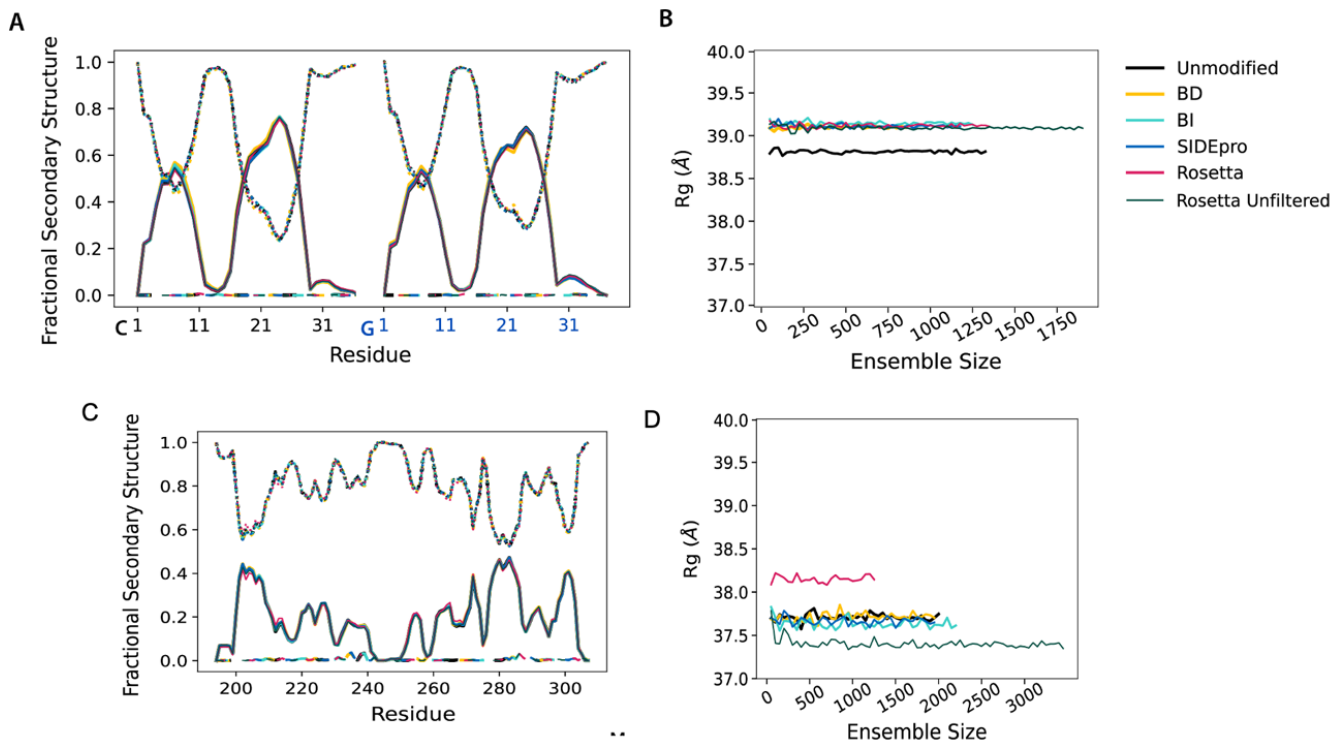

**Figure S7**. *Histone H3 and UDFI conformers generated with different PTMs libraries.* Fractional secondary structure propensities for (A) histone H3 N-terminal IDRs on chain C and G and (C) UDFI. Secondary structures were calculated using DSSP. Helices are shown in solid line, loop in dotted line and β-sheet in loosely dashed line. Mean radius of gyration over ensemble size, calculated from an average of 30 randomly selected ensemble for (B) histone H3 N-terminal IDRs on chain C and G and (D) UDFI

Reference

1    King, J. E. & Koes D. V. Sidechainnet: An all-atom protein structure dataset for machine learning. *Proteins: Structure, Function, and Bioinformatics* **89**, 1489–1496 (2021).